\documentclass[journal ]{new-aiaa}
\usepackage[utf8]{inputenc}
\usepackage{textcomp}

\usepackage{graphicx}
\usepackage{subcaption}
\usepackage{amsmath}
\usepackage[version=4]{mhchem}
\usepackage{siunitx}
\usepackage{longtable,tabularx}
\title{Characterizing 3D Reattached Flow on an Airfoil with Finite-Span Synthetic Jets}

\author{Adnan Machado \footnote{MASc Student, Mechanical and Industrial Engineering, 5 King's College Rd, Toronto, Ontario, Canada} and Pierre E. Sullivan \footnote{Professor, Mechanical and Industrial Engineering, 5 King's College Rd, Toronto, Ontario, Canada}}
\affil{University of Toronto, Toronto, Ontario, M5S3H2}

\begin{document}

\maketitle


\section*{Nomenclature}
{\renewcommand\arraystretch{1.0}
\noindent\begin{longtable*}{@{}l @{\quad=\quad} l@{}}
$b$ & Wing span [\unit{\milli\metre}]\\
$b_a(x)$ & Effective spanwise control length [\unit{\milli\metre}]\\
$b_{\mathrm{SJA}}$ & Span of synthetic jet actuator array [\unit{\milli\metre}]\\
$c$ & Airfoil chord length [\unit{\milli\metre}]\\
$\mathrm{Re}_c$ & chord-based Reynolds number\\
$U_\infty$ & Freestream velocity [\unit[per-mode=symbol]{\metre\per\second}]\\
$x$, $y$, $z$ & Chordwise, transverse, and spanwise coordinate [\unit{\milli\metre}]\\
$x_s(z)$ & Chordwise separation point [\unit{\milli\metre}]\\

$\alpha$ & Angle of attack [\unit{\degree}] \\

\end{longtable*}}

\section{Introduction}
Active flow control systems with synthetic jet actuators (SJAs) can improve airfoil performance at low Reynolds numbers by mitigating stall. Their lightweight, power-efficient design and ease of integration make them well suited for applications in aviation and wind turbines. These systems have achieved a high technology readiness level (TRL) of 7, as demonstrated by successful flight test implementations~\cite{Seifert2010,Greenblatt2022}. Synthetic jets reattach separated flows by enhancing mixing between the freestream and separated shear layer~\cite{Salunkhe2016,Liu2022}. Active flow control systems using SJAs are commonly implemented in two configurations: finite-span slot-style SJAs or arrays of small-orifice discrete SJAs, both typically shorter than the span of the wing being controlled, resulting in three-dimensional flows~\cite{Sahni2011,Feero2017a}. Prior experiments have shown that even at large momentum coefficients, that the effective spanwise control length was limited to approximately 40\% of the SJA array length~\cite{Machado2024b,Machado2024c}.

While active flow control with synthetic jets has been extensively studied, most investigations have focused on midspan measurements, overlooking the inherently three-dimensional nature of flows reattached by finite-span actuators. As a result, while significant gains have been achieved in the midspan lift coefficient, great potential remains to further enhance overall aerodynamic performance by expanding the spanwise control authority. Realizing this potential requires a comprehensive understanding of how reattached flows develop along the span to optimize control effectiveness and improve overall lift.

To address this gap, previous studies investigated the three-dimensional flow development over a stalled airfoil controlled by an array of circular SJAs. These efforts included flow visualizations across multiple spanwise planes, supported by quantitative measurements of the mean flow and unsteady dynamics~\cite{Machado2024a,Machado2024b, FeeroPhD}. Additionally, orthogonal smoke flow visualizations in the streamwise-spanwise and spanwise-transverse planes were conducted with a horizontal smoke wire, contributing to a more comprehensive understanding of the flow~\cite{Machado2024c}. Much of the initial understanding of the spanwise behavior emerged through repeated naked-eye observations, which revealed consistent flow characteristics across a wide range of control parameters, including blowing strength, excitation frequency, and duty cycle. These observations guided targeted measurements to confirm and detail the spatial flow patterns in all three orthogonal planes.

This technical note synthesizes the findings discussed above and incorporates new visualizations to develop a general, physics-based description of flows reattached by finite-span SJA arrays. The controlled flow exhibits three distinct regions: a directly controlled region, an unaffected region, and a transitional zone between them. By integrating measurements and visualizations from both discrete and slot-style SJA configurations, this work provides general insights into the three-dimensional shear layer behavior. Lastly, a parameter to evaluate the spanwise control authority of an SJA array is presented, providing a practical metric for evaluating and guiding future actuator designs.

\section{Experimental Method}
Wind tunnel tests were conducted on a NACA 0025 airfoil in the low-speed recirculating wind tunnel in the Department of Mechanical and Industrial Engineering at the University of Toronto. The aluminum wing has an aspect ratio of approximately 3, with a span of $b=885$~\unit{\milli\metre}, and a chord length of $c=300$~\unit{\milli\metre}. The wind tunnel was operated at a freestream velocity of $U_\infty=5.1$~\unit[per-mode = symbol]{\metre\per\second}, resulting in a chord-based Reynolds number of $\mathrm{Re}_c=10^5$. With the angle of attack set to $\alpha=10$~\unit{\degree}, the flow separates without reattachment, leaving the airfoil in a stalled state.

An array of 12 Murata MZB1001T02 microblower SJAs, located at 10.7\% chord and spaced \SI{25}{\milli\metre} apart, is installed at the center of the wing, for a total effective span of $b_{\mathrm{SJA}}=275$~\unit{\milli\metre}. The SJAs ingest and expel fluid at a high frequency through \SI{0.8}{\milli\metre} nozzles and are flush-mounted beneath the airfoil surface. Prior tests have demonstrated their effectiveness in flow separation control~\cite{Machado2024a,Machado2024b}.

\begin{figure}
    \centering
    \begin{subfigure}{0.3459\textwidth}
    \centering
        \includegraphics[width=\linewidth]{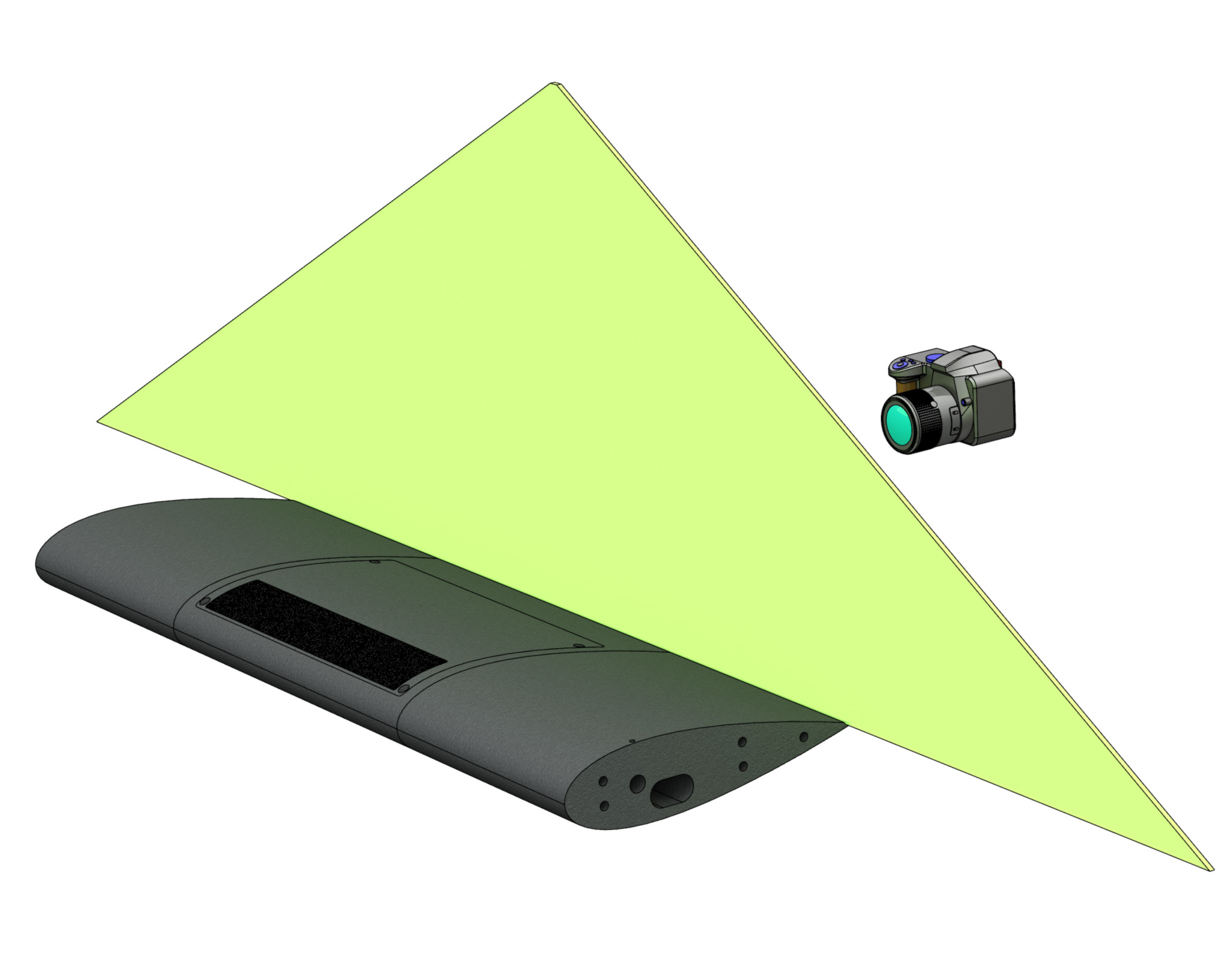}
        \caption{Cross-sectional with laser sheet}
        \label{fig:sectional}
    \end{subfigure}
    \begin{subfigure}{0.32\textwidth}
    \centering
        \includegraphics[width=\linewidth]{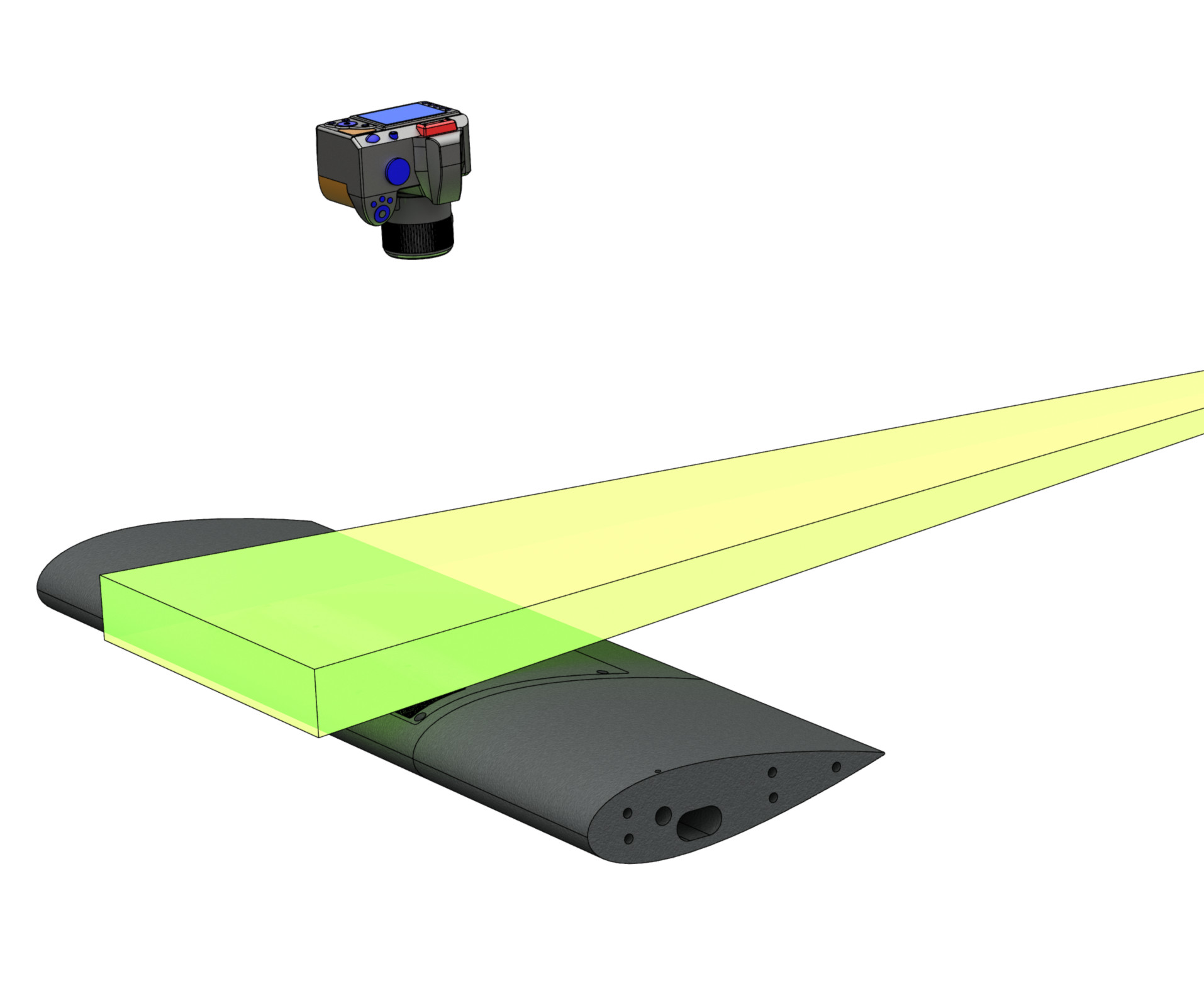}
        \caption{Overhead with laser volume}
        \label{fig:overhead}
    \end{subfigure}
    \begin{subfigure}{0.32\textwidth}
    \centering
        \includegraphics[width=\linewidth]{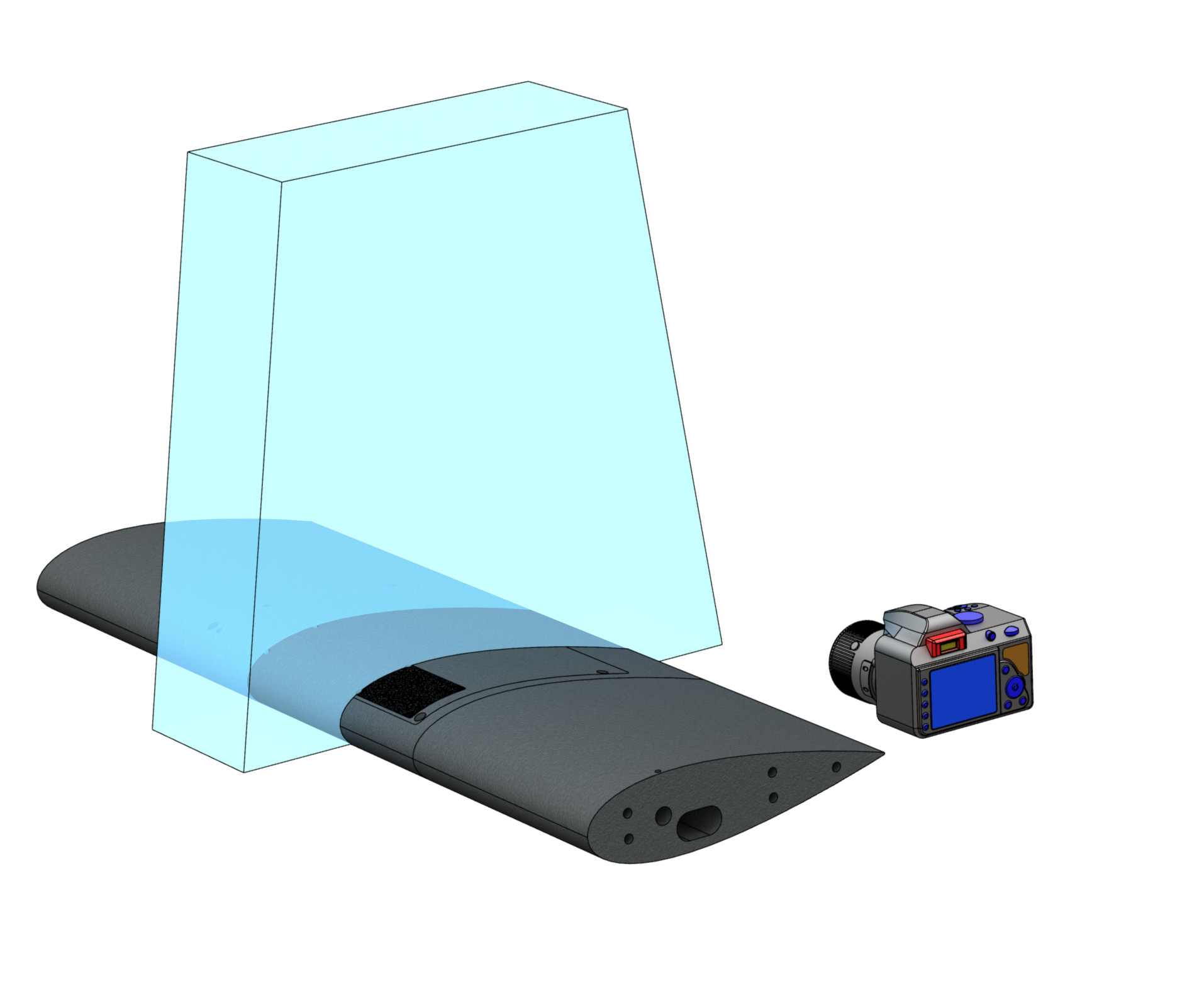}
        \caption{Side-view with speedlight}
        \label{fig:side}
    \end{subfigure}
    \caption{Camera and illumination configurations used for the smoke flow visualizations}
    \label{fig:methods}
\end{figure}

Smoke visualizations were taken at three orthogonal planes, as illustrated in Figure~\ref{fig:methods}. A horizontal smoke wire, placed just upstream of the airfoil's stagnation point, generated streaklines that traced the shear layer boundary for the configurations shown in Figures~\ref{fig:sectional} and \ref{fig:overhead}. A green laser, modified with optics to provide precise illumination, maximized contrast by ensuring that the background remained dark. For the more conventional side-view visualizations, a Nikon SB-800 speedlight was used to illuminate the smoke streams, as shown in Figure~\ref{fig:side}. The three configurations enable a detailed visualization of the flow components in the $x$, $y$, and $z$ directions, offering a comprehensive, three-dimensional perspective of the mean flow field. Images were captured using a Nikon D7000 DSLR camera with a remote shutter. Further details on the smoke visualization technique can be found in \citet{Machado2024a,Machado2024b}.

For comparison, this paper presents mean velocity fields obtained with particle image velocimetry by \citet{FeeroPhD}. Similar experimental conditions were used with the same wing model; however, a rectangular slot-style SJA, spanning $b_{\mathrm{SJA}}=$~\SI{294}{\milli\metre} and centered at midspan, was used instead. The angle of attack was set to $\alpha=12$~\unit{\degree}, and the Reynolds number was $\mathrm{Re}_c=1.25 \times 10^5$.

\section{Discussion}
\begin{figure}
    \centering
    \includegraphics[width=0.8\linewidth]{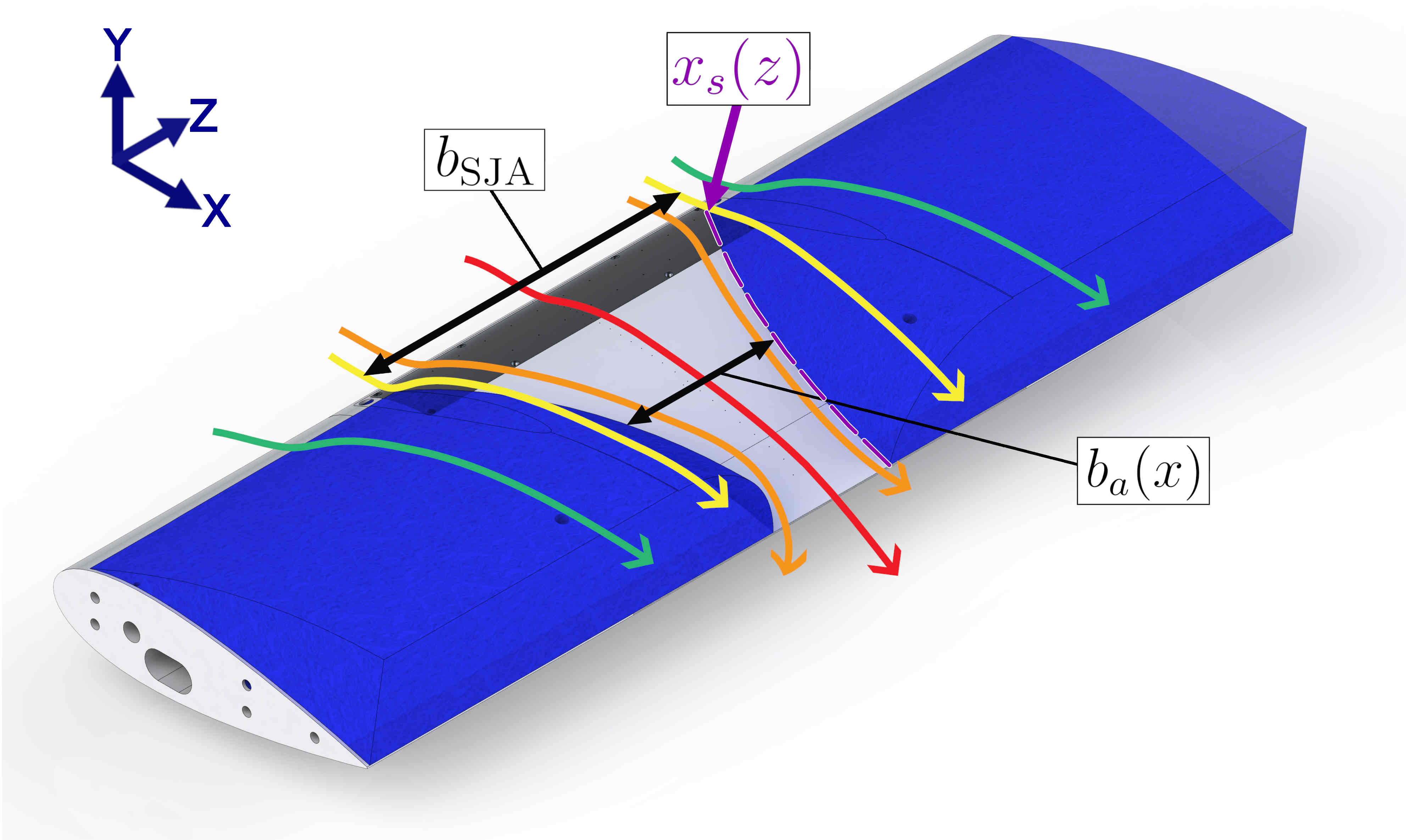}
    \caption{Isometric view of the conceptualized flow with depiction of the oncoming flow}
    \label{fig:cartoon-isometric}
\end{figure}

Figure~\ref{fig:cartoon-isometric} provides an isometric view of the conceptualized flow, with the blue volume over the airfoil representing the turbulent, high-pressure recirculation region, while the absence of blue denotes the attached flow region. Additionally, three key parameters are defined: $b_a(x)$, representing the span of the attached flow region at a given chordwise location, $b_{\mathrm{SJA}}$, representing the spanwise length of the SJA array, and $x_s(z)$, denoting the chordwise separation point at a specific spanwise position.

\begin{figure}
    \centering
    \begin{subfigure}{0.445\textwidth}
    \centering
        \includegraphics[width=\linewidth]{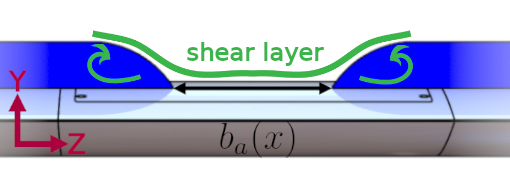}
        \caption{}
        \label{fig:cartoon_sec_60c}
    \end{subfigure}
    \begin{subfigure}{0.445\textwidth}
    \centering
        \includegraphics[width=\linewidth]{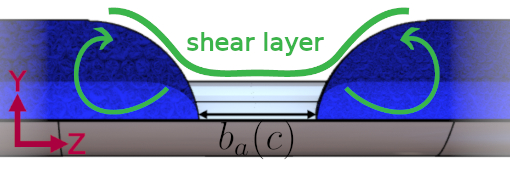}
        \caption{}
        \label{fig:cartoon_sec_trailing_edge}
    \end{subfigure}
    \begin{subfigure}{0.445\textwidth}
    \centering
        \includegraphics[width=\linewidth]{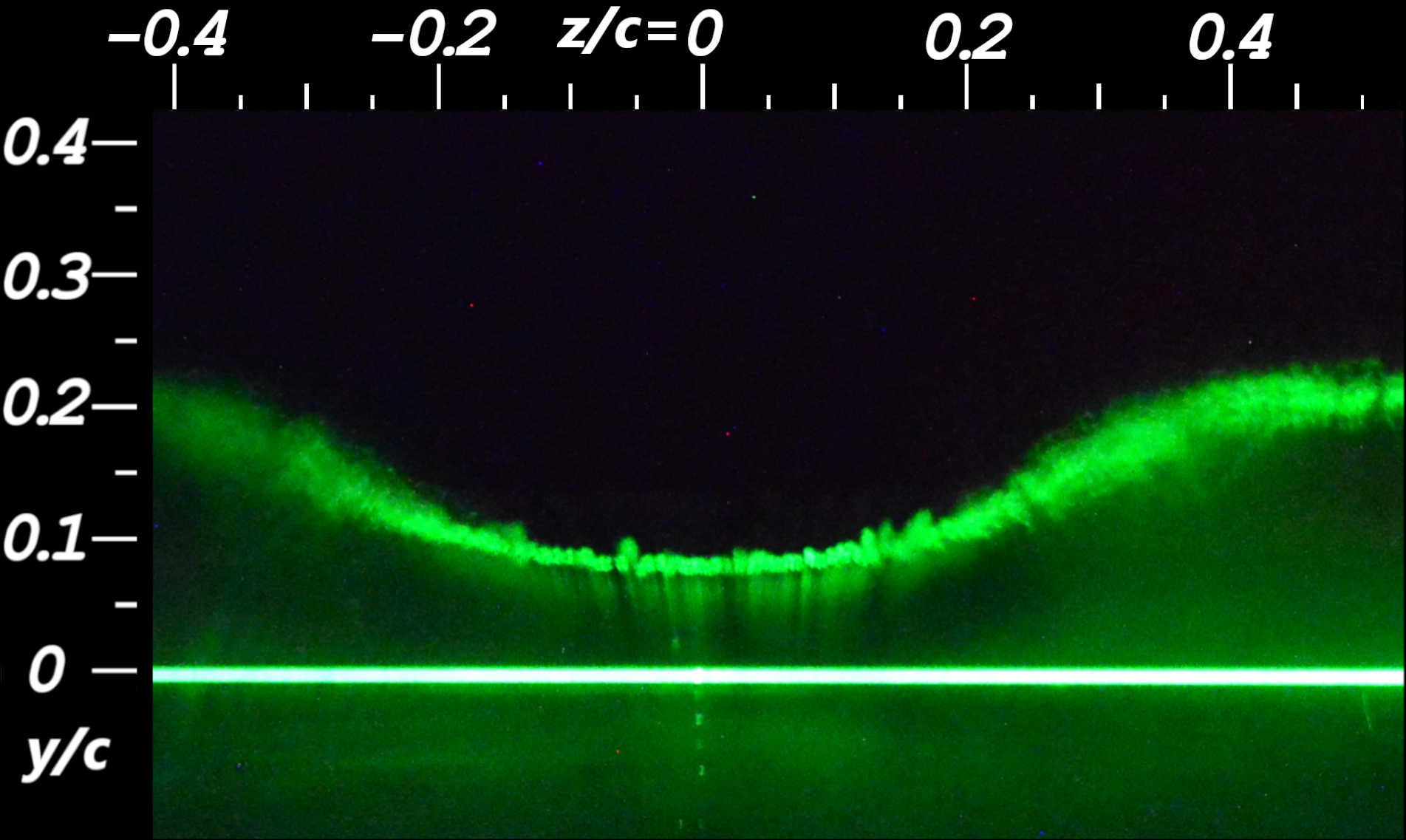}
        \caption{}
        \label{fig:smoke_cartoon_sec_60c}
    \end{subfigure}
    \begin{subfigure}{0.445\textwidth}
    \centering
        \includegraphics[width=\linewidth]{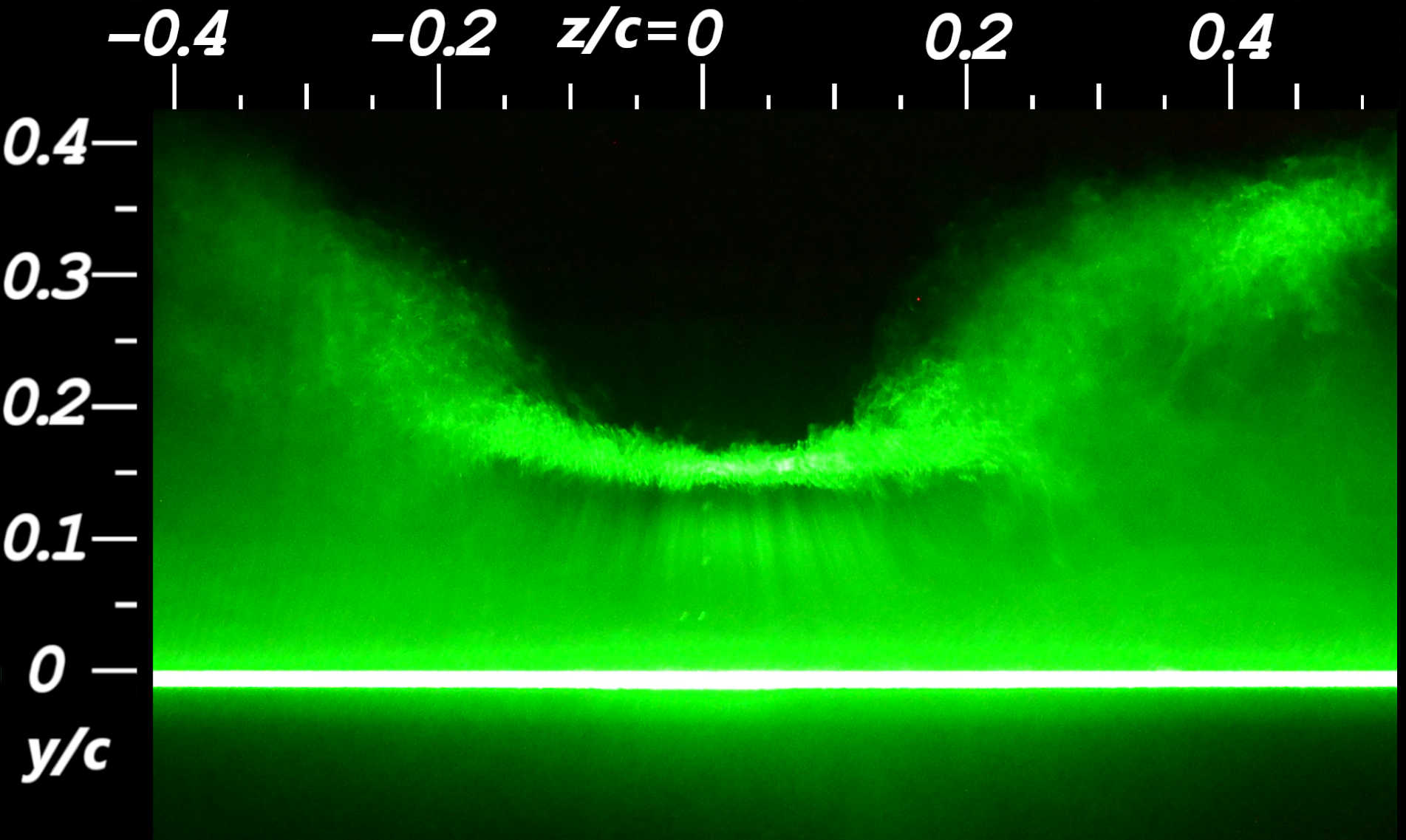}
        \caption{}
        \label{fig:smoke_cartoon_sec_trailing_edge}
    \end{subfigure}
    \caption{Conceptualized flow (a, b) compared with smoke visualizations of the shear layer (c, d) for cross-sectional views at 0.6$c$ (a, c) and the trailing edge (b, d)}
    \label{fig:cartoon_comparison_sectional}
\end{figure}

\begin{figure}
    \centering
    \begin{subfigure}{\textwidth}
    \centering
        \includegraphics[width=0.8\linewidth]{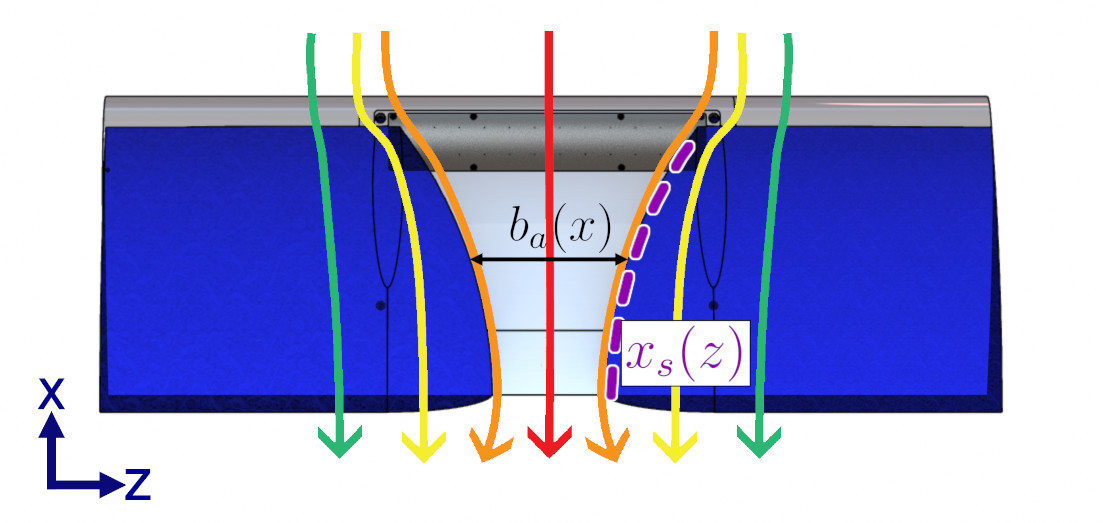}
        \caption{}
        \label{fig:cartoon_overhead}
    \end{subfigure}
    \newline
    \begin{subfigure}{0.5\textwidth}
    \centering
        \includegraphics[width=\linewidth]{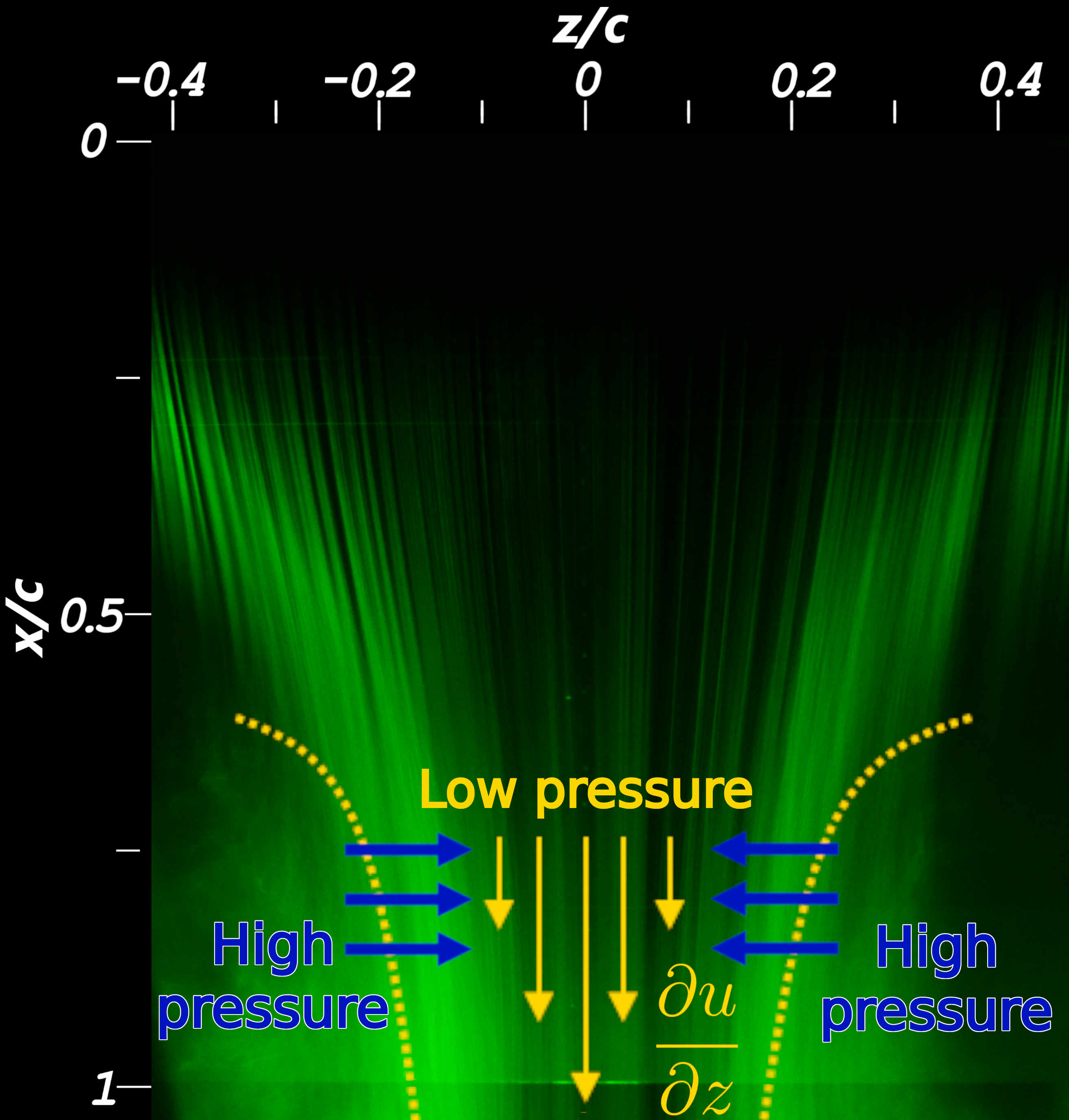}
        \caption{}
        \label{fig:smoke_cartoon_overhead}
    \end{subfigure}
    \caption{Synthesized flow (a) compared with smoke visualization (b) for the overhead view}
    \label{fig:cartoon_comparison_overhead}
\end{figure}

\begin{figure}
    \centering
    \includegraphics[width=0.4\linewidth]{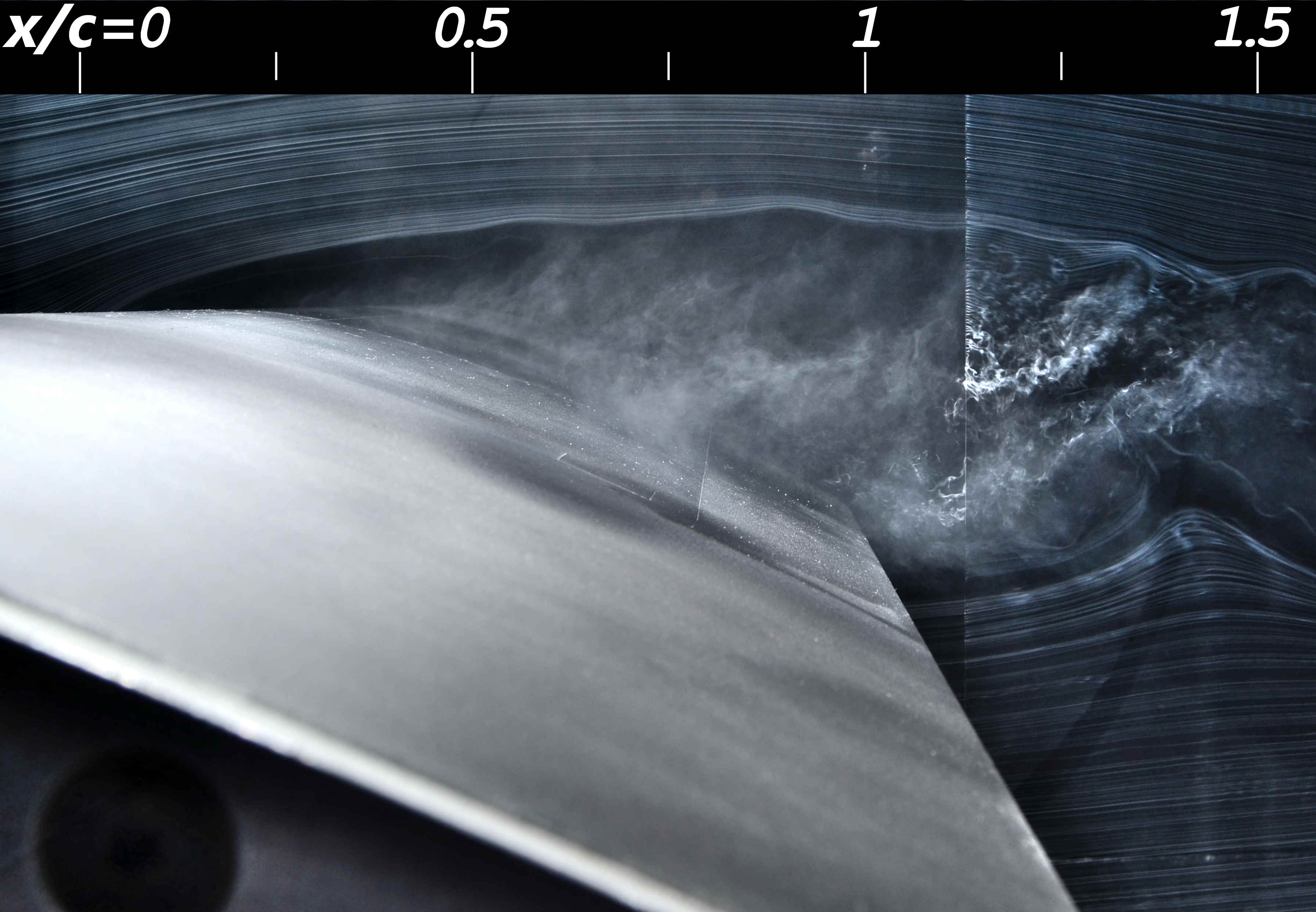}
    \caption{Smoke visualization of the baseline flow}
    \label{fig:baseline}
\end{figure}

\begin{figure}
    \centering
    \begin{subfigure}{0.3\textwidth}
        \includegraphics[width=\linewidth]{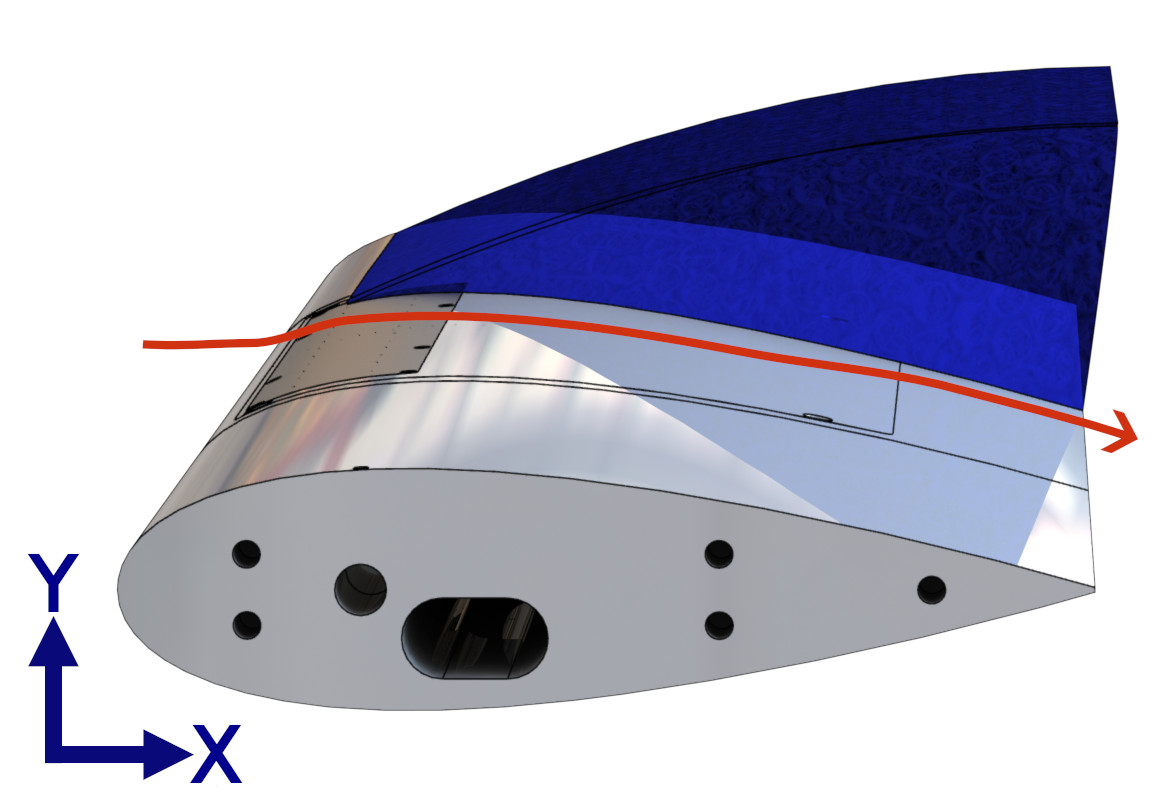}
        \caption{}
        \label{fig:cartoon_midspan}
    \end{subfigure}
    \begin{subfigure}{0.3\textwidth}
        \includegraphics[width=\linewidth]{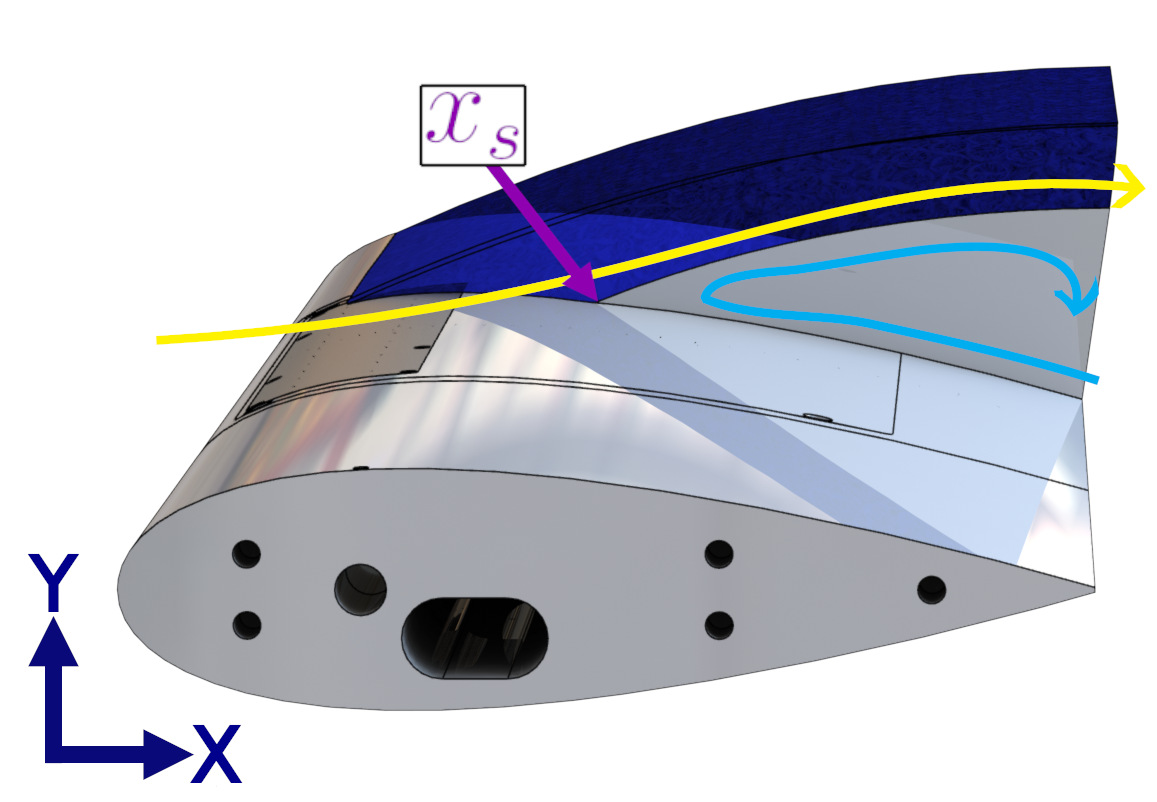}
        \caption{}
        \label{fig:cartoon_intermediate}
    \end{subfigure}
    \begin{subfigure}{0.3\textwidth}
        \includegraphics[width=\linewidth]{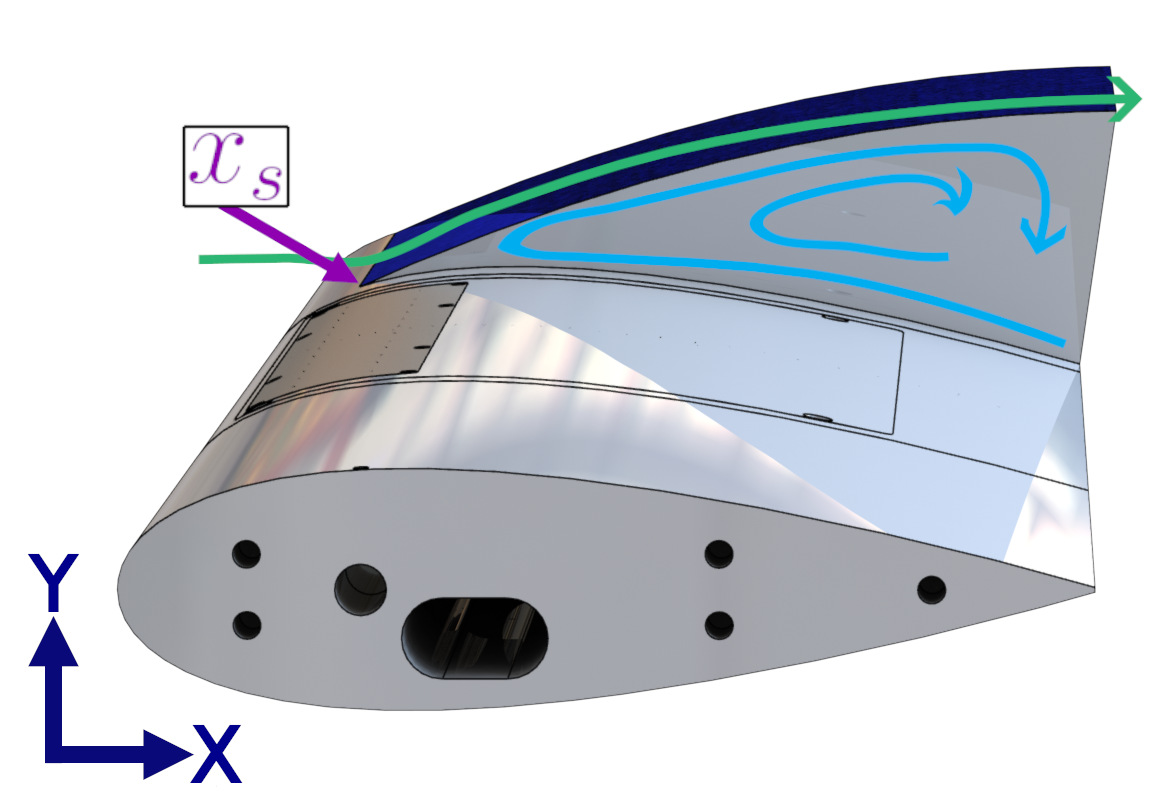}
        \caption{}
        \label{fig:cartoon_baseline}
    \end{subfigure}
    \begin{subfigure}{0.3\textwidth}
        \includegraphics[width=\linewidth]{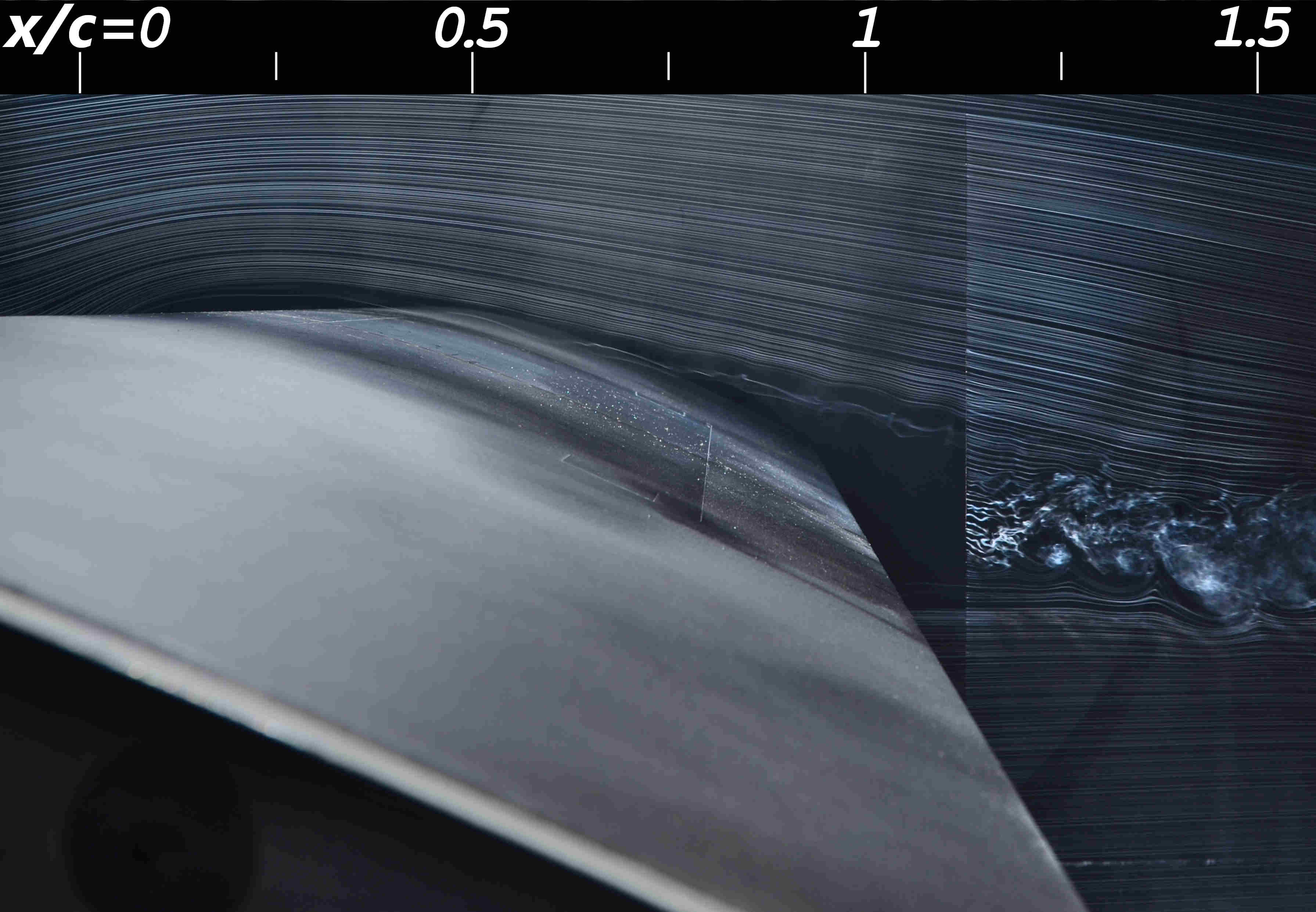}
        \caption{}
        \label{fig:smoke_cartoon_midspan}
    \end{subfigure}
    \begin{subfigure}{0.3\textwidth}
        \includegraphics[width=\linewidth]{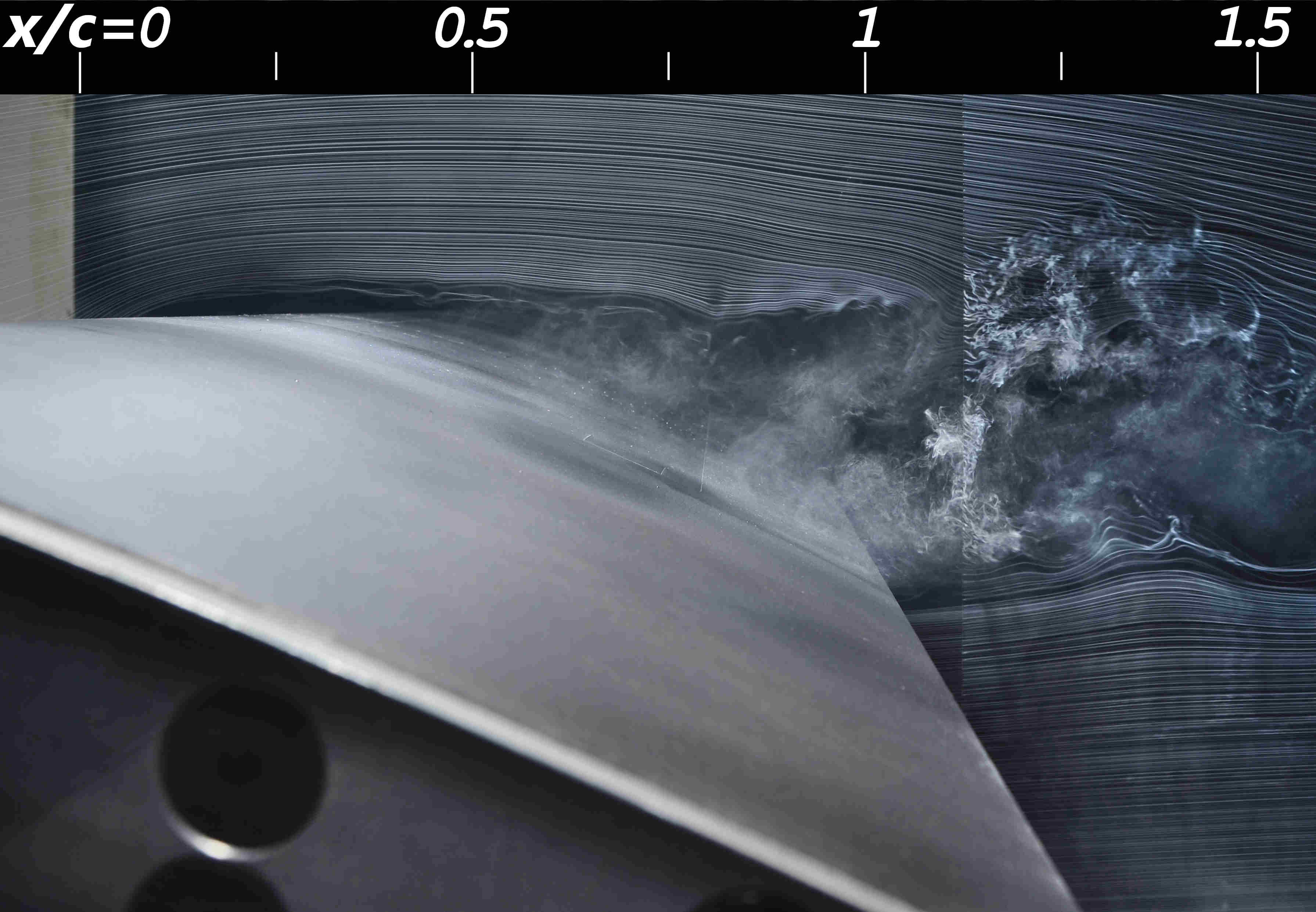}
        \caption{}
        \label{fig:smoke_cartoon_intermediate}
    \end{subfigure}
    \begin{subfigure}{0.3\textwidth}
        \includegraphics[width=\linewidth]{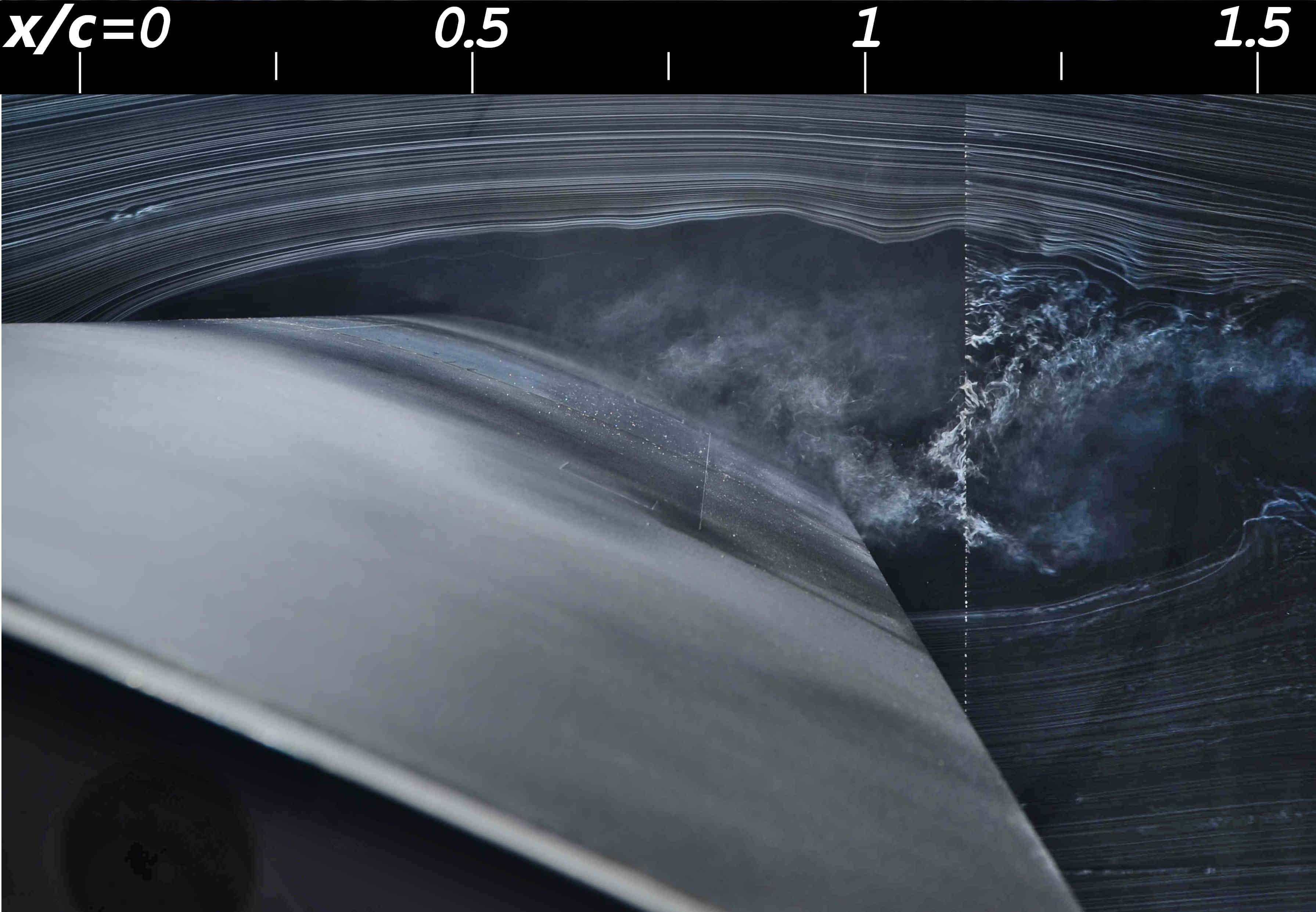}
        \caption{}
        \label{fig:smoke_cartoon_baseline}
    \end{subfigure}
    \begin{subfigure}{0.3\textwidth}
        \includegraphics[width=\linewidth]{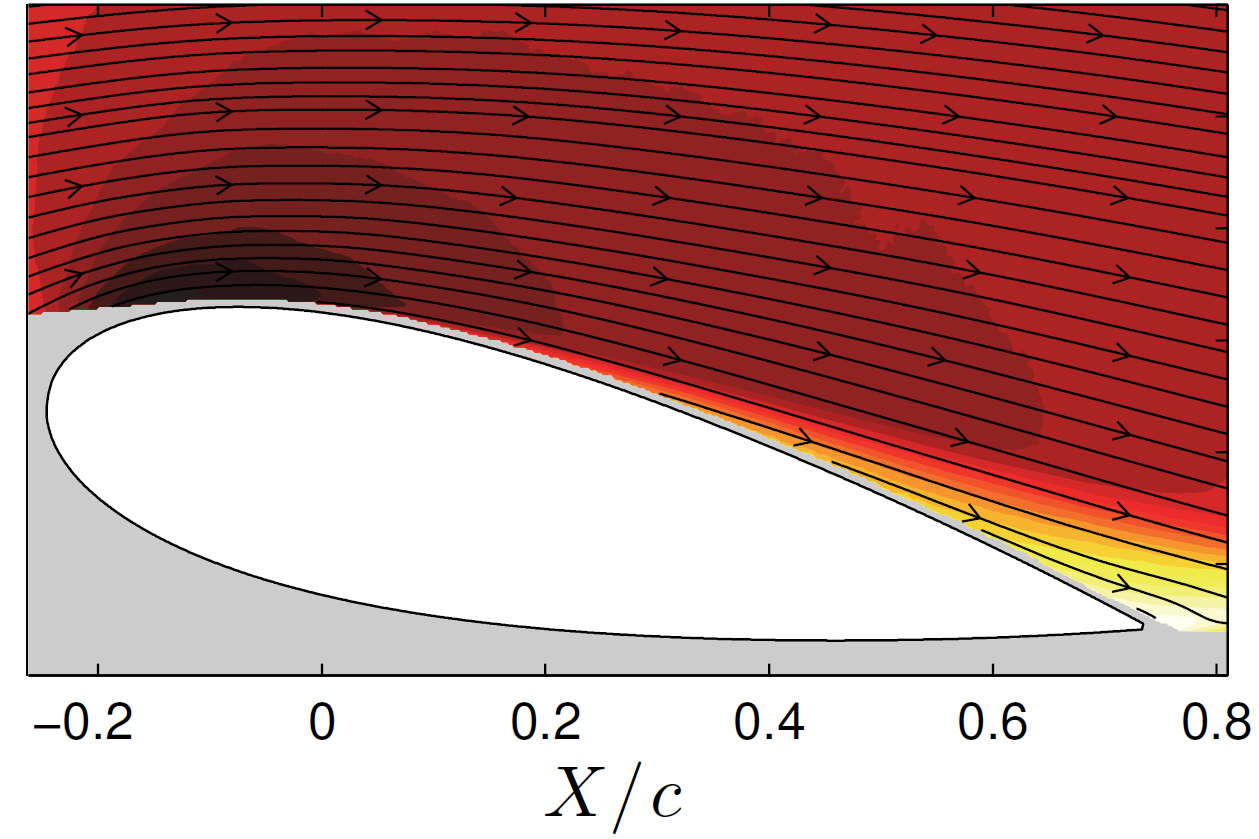}
        \caption{}
        \label{fig:feero_midspan}
    \end{subfigure}
    \begin{subfigure}{0.3\textwidth}
        \includegraphics[width=\linewidth]{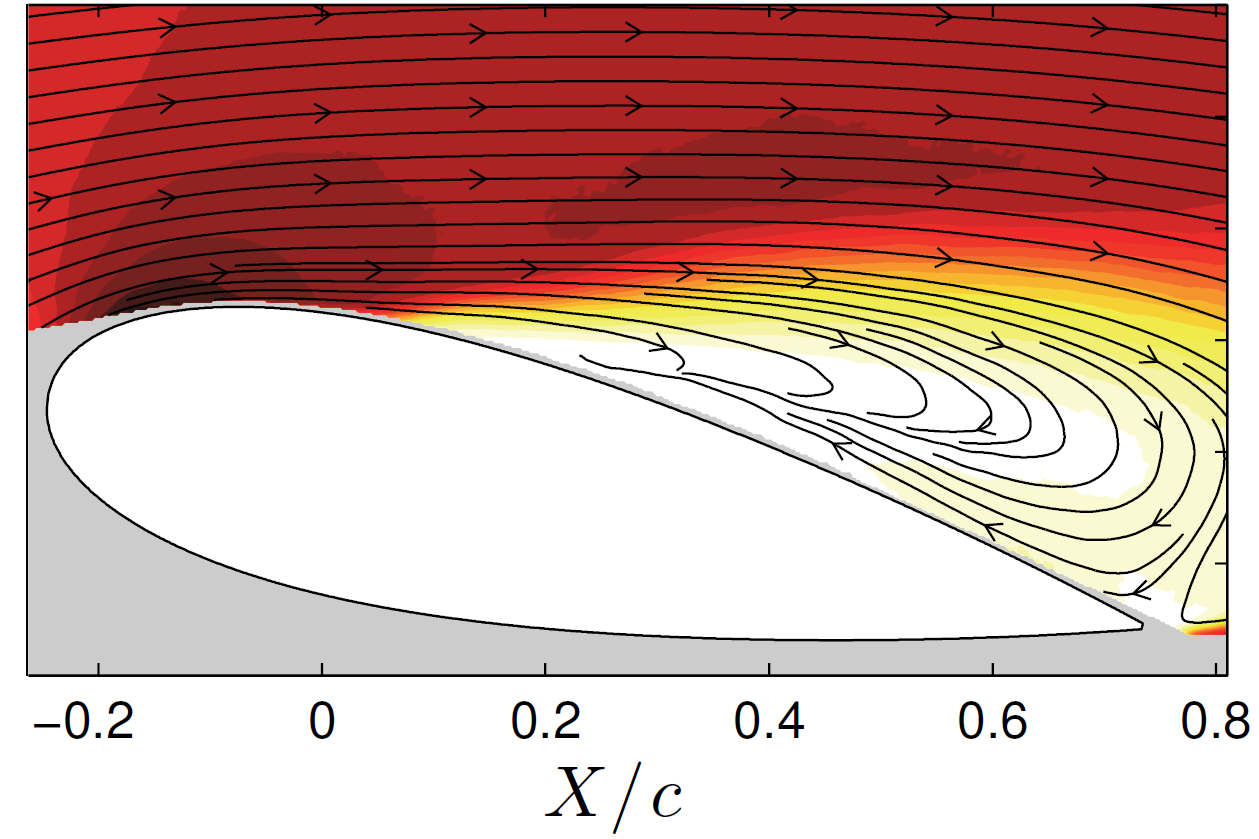}
        \caption{}
        \label{fig:feero_middle}
    \end{subfigure}
    \begin{subfigure}{0.3\textwidth}
        \includegraphics[width=\linewidth]{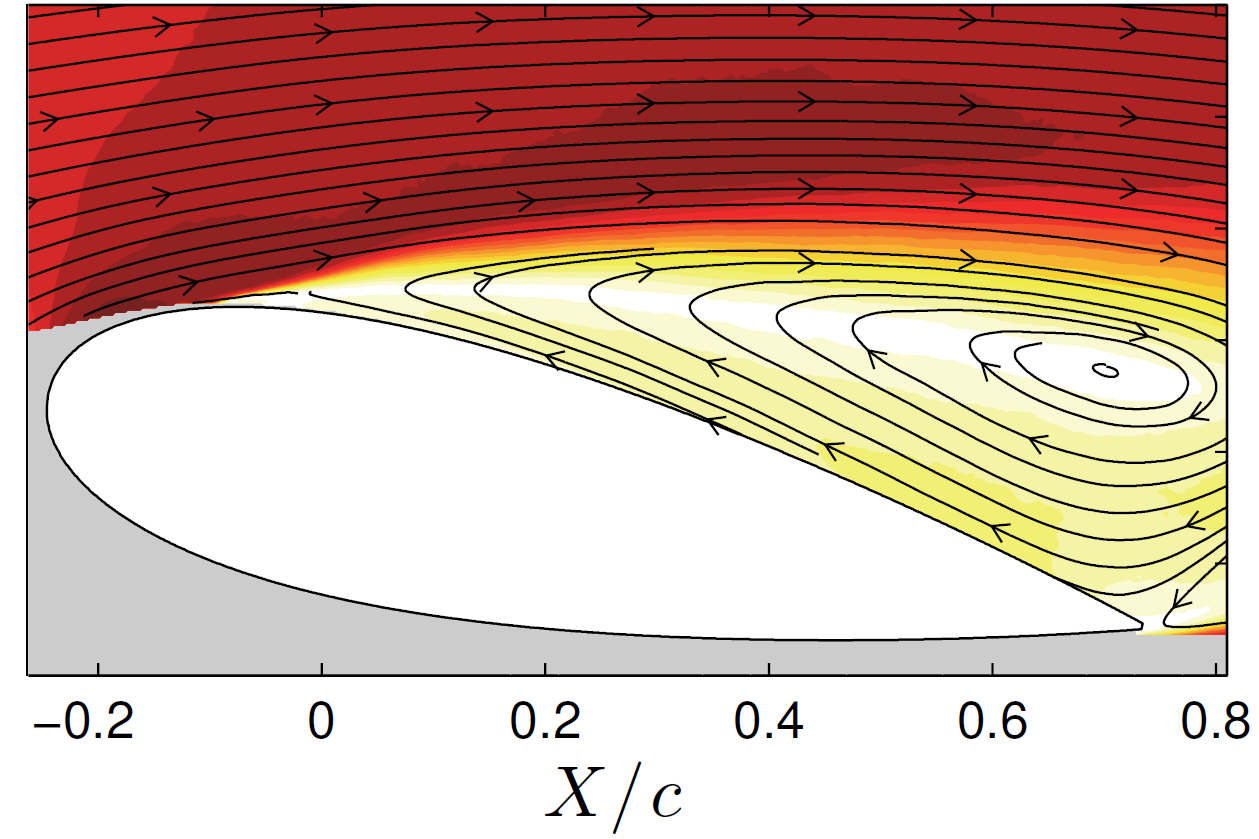}
        \caption{}
        \label{fig:feero_baseline}
    \end{subfigure}
    \caption{Flow structure representation (a–-c), compared with smoke visualizations (d–-f), and PIV velocity fields (g-–i) adapted from \citet{FeeroPhD} with permission. The flow is shown at the midspan (a, d, g), an intermediate spanwise distance (b, e, h), and a far spanwise distance (c, f, i)}
    \label{fig:cartoons_comparison_vertical}
\end{figure}

The reattached flow over the airfoil is modeled as three distinct control regions:
\begin{itemize}
    \item \textbf{Controlled Region:} Flow separation is suppressed over a spanwise length, $b_a(x)$, centered around the midspan. This effective spanwise control length decreases monotonically along the chord.
    \begin{itemize}
        \item Figure~\ref{fig:cartoon_comparison_sectional} provides cross-sections of the shear layer in the $y$-$z$ plane, illustrating the spanwise length of the reattached flow at different chordwise positions. At the midchord (Figure~\ref{fig:cartoon_sec_60c}), the shear layer exhibits a longer spanwise attached length compared to the trailing edge (Figure~\ref{fig:cartoon_sec_trailing_edge}). This model is validated through comparisons with experimental smoke flow visualizations of the shear layer cross-section. Figure~\ref{fig:smoke_cartoon_sec_60c} presents a visualization of the shear layer at $x/c=0.6$, where a large attached region -- marked by dense, illuminated smoke -- is observed, along with a shallow curvature as the shear layer thickens along the span. In contrast, at the trailing edge (Figure~\ref{fig:smoke_cartoon_sec_trailing_edge}), the attached region is seen to be much shorter. Beyond this attached region, recirculating turbulent flow is evident from the diffuse smoke spreading in the $y$-direction.
        \item The overhead view in Figure~\ref{fig:cartoon_comparison_overhead} complements this understanding of the flow field. The flow exhibits a spanwise contraction which is depicted in Figure~\ref{fig:cartoon_overhead}. This contraction results in the decrease of the spanwise control length $b_a$, along the chord. Figure~\ref{fig:smoke_cartoon_overhead} presents a visualization of the controlled flow where smooth smoke streaks indicate laminar flow. Further downstream, the recirculation region grows, forcing the flow inward towards the midspan. The spanwise velocity gradient, $\frac{\partial u}{\partial z}$, induces a spanwise pressure gradient leading to the observed contraction in the smoke streaks. A similar spanwise flow contraction was observed in prior studies by \citet{Sahni2011,Feero2017a}.
        \item The spanwise control authority of an SJA, a crucial parameter for evaluating effectiveness, can be quantified by the ratio of attached flow at the trailing edge to the span of the SJA array, i.e. $\frac{b_a(c)}{b_{\mathrm{SJA}}}$. A higher ratio value signifies greater spanwise control authority, indicating that the actuator is effective at suppressing flow separation across a larger spanwise region.
    \end{itemize}
    \item \textbf{Uncontrolled Region:} Beyond the effective span of the SJA array, the flow reverts to baseline conditions, exhibiting a large recirculation region depicted by the outer blue region beyond the curve in Figures~\ref{fig:cartoon_sec_60c} and \ref{fig:cartoon_sec_trailing_edge}. In this region, the flow is unaffected by the SJAs and resembles the fully separated flow of the baseline case, as shown in Figure~\ref{fig:baseline}. This effect is most clearly seen in the side view depiction, smoke visualization, and the velocity field presented in Figures~\ref{fig:cartoon_baseline}, \ref{fig:smoke_cartoon_baseline}, and \ref{fig:feero_baseline}, respectively. \pagebreak
    \item \textbf{Transitional Region:} As the flow transitions along the span from fully attached to fully separated, the recirculation region forms a distinctive curved shape in both the $x$-$z$ and $y$-$z$ planes, encroaching toward the midspan as the flow progresses along the chord.
    \begin{itemize}
        \item This behavior is best understood through multiple 2D flow slices, as shown in Figure~\ref{fig:cartoons_comparison_vertical}. Near the midspan, the flow remains fully attached (Figure~\ref{fig:cartoon_midspan}). Moving away from the midspan (Figure~\ref{fig:cartoon_intermediate}), the control effects diminish, leading to a recirculation zone at the trailing edge. Flow separation occurs, with the chordwise separation point varying along the span. As the spanwise distance increases (increasing $|z|$), the separation point, $x_s$, shifts upstream, and the recirculation area expands upward, forcing the incoming flow onto a higher trajectory. Far from the midspan (Figure~\ref{fig:cartoon_baseline}), the control effects become negligible, and the recirculation region resembles the baseline case, with a separation point that matches the baseline flow. Experimental support for this model comes from smoke visualizations (Figures~\ref{fig:smoke_cartoon_midspan} to \ref{fig:smoke_cartoon_baseline}) and PIV velocity fields (Figures~\ref{fig:feero_midspan} to \ref{fig:feero_baseline}).
        \item The expansion of the recirculation region is seen in the sectional flow visualization in Figure~\ref{fig:smoke_cartoon_sec_trailing_edge}, where diffuse smoke rises in the $y$-direction along a curved path. This shape, depicted in Figure~\ref{fig:cartoon_sec_trailing_edge} and, along with the overhead perspective in Figure~\ref{fig:cartoon_overhead}, provides a comprehensive view of how recirculation initiates as a localized trailing-edge region and then expands upward and upstream with increasing spanwise distance.
    \end{itemize}
\end{itemize}

\newpage
\section{Conclusion}
This paper presents a synthesized, three-dimensional description of reattached flow over an airfoil controlled by a finite-span SJA array. Orthogonal smoke visualizations and PIV data are integrated to characterize the flow by three distinct regions: directly controlled, transitional, and unaffected. The transitional region is of particular interest, marking the shift from controlled to uncontrolled flow, exhibits a curved recirculation boundary that encroaches toward the midspan. Lastly, a parameter, $\frac{b_a(c)}{b_{\mathrm{SJA}}}$, for evaluating spanwise control authority is discussed, offering a framework for evaluating finite-span active flow control systems.

This flow characterization improves understanding and guides targeted measurements to optimize active flow control system design for lift enhancement on wings and turbine blades. These insights will facilitate the transition of synthetic jet actuation toward real-world applications, ultimately enabling expanded flight envelopes and improved aerodynamic efficiency.
\clearpage

\section*{Funding Sources}
This work was supported by the Natural Science and Engineering Research Council of Canada and the SciNet High Performance Computing consortium and the Canadian Microelectronics Corporation (CMC).

\bibliography{sample}

\end{document}